%
%
%
%
%
\input amstex
\documentstyle{amsppt}
\Monograph
\loadbold

\NoRunningHeads
\catcode`\@=11
\def\logo@{}
\def\proclaimheadfont@{\smc}
\def\demoheadfont@{\smc}

\def\output@{\shipout\vbox{%
 \iffirstpage@ \global\firstpage@false
   \hbox to\hsize
    {\hfil\sixrm 
     UNIVERSITATIS IAGELLONICAE ACTA MATHEMATICA, FASCICULUS XXXIV
     \hfil}
   \vskip 2pt
   \hrule
  \vskip 4pt 
  \hbox to\hsize{\sixrm\hfil 1996 \hfil}
  \pagebody 
 \else
     \hbox to\hsize{{\tenrm\ifodd\pageno \hss\folio \else \folio\hss \fi}}
   \vskip 0.7truecm
%
   \pagebody
 \fi}%
 \advancepageno \ifnum\outputpenalty>-\@MM\else\dosupereject\fi}

\catcode`\@=\active

\magnification=\magstephalf

\baselineskip 18pt

\hsize 13 truecm
\vsize 19 truecm

\LimitsOnSums
\LimitsOnNames
\CenteredTagsOnSplits

\def\rec#1{{\eightpoint{\it Received #1}\hfill\newline}}
\def\addr#1{{\eightpoint\phantom{.}\hskip 7.5truecm#1 \newline}}

\def\theo#1{\procla{Theorem #1} }
\let\endtheo\endproclaim

\def\lem#1{\procla{Lemma #1} }
\let\endlem\endproclaim

\def\proof#1{\demol{Proof #1}}
\let\endproof\enddemo

\def\prop#1{\procla{Proposition #1}}
\let\endprop\endproclaim

\def\cor#1{\procla{Corollary #1}}
\let\endcor\endproclaim

\def\defi#1{\demol\nofrills{Definition #1.\usualspace}}
\let\enddefi\enddemo

\def\rem#1{\demol\nofrills{Remark #1.\usualspace}}

\let\endrem\enddemo

\let\demol\demo
\let\procla\proclaim

\def\longmapsto{\DOTSB\mapstochar\longrightarrow}

\define \bo {\bold}
\define \oti {\otimes}

\define\pc#1#2{ \frac {\partial {#1}}{\partial {#2}} }
\define\pcc#1#2{ \frac {\eth {#1}}{\eth {#2}} }
\define\pct#1#2#3{ \frac {\partial^{#1} {#2}}{\partial {#3}^{#1}} }
\define\pzt#1#2#3{ \frac {d^{#1} {#2}}{d{#3}^{#1}} }
\define\pz#1#2{ \frac {d{#1}}{d{#2}} }

%
%
%
%
%
\loadeusm 
\def\proof#1{\demol{\eightpoint Proof #1}\eightpoint} 
\let\endproof\enddemo                                 

\def\sproof#1{\demol{\eightpoint Sketch of Proof #1}\eightpoint}
\let\endsproof\enddemo
%
%
\def\POLG{\mathord{{\eusm L}_{\!+}^{\!\uparrow}}}
\def\PO{\mathord{\widetilde{\eusm P}}}
\def\MM{\mathord{\eusm M}}
\def\spec{\mathord{\roman{spec}}}
\def\forw{\mathord{\overline{V}^{\,+}}}
\def\SM{\mathord{\eusm S(\eusm M)}}
\def\SMi{\mathord{\eusm S(\eusm M^i)}}
\def\SMm{\mathord{\eusm S(\eusm M^m)}}
\def\SMn{\mathord{\eusm S(\eusm M^n)}}
\def\SB{\mathord{\underline{\eusm S}}}
\def\SRR{\mathord{\eusm S(\bold R^2)}}
\def\Dom{\mathord{\eusm D}}
\def\Do{\mathord{\Dom_0}}
\def\Falg{\mathord{\eusm F}}
\def\Palg#1{\mathord{\eusm P \left(\left\{ #1 \right\}\right)}}
\def\komm#1#2{\mathord{\left[#1\mathord,#2\right]}}
\def\supp{\mathord{\roman{supp}}}
\def\WW{\mathord{\frak W}}
\def\Wf{\mathord{\bold W}}
\def\SPACE#1{\frak #1} 
\def\sV{\SPACE{V}}
\def\VV{\sV}
\def\VN{{\sV}_N}
\def\VND{{\sV}_N^\ast}
\def\Vp{{\sV}^+}
\def\Vpp{{\sV}^{++}}
\def\Vm{{\sV}^-}
\def\Vmm{{\sV}^{--}}
\def\KK{\SPACE{K}}
\def\HH{\SPACE{H}}
\def\HPH{\mathord{\SPACE{H}_{\roman{phys}}}}
\def\SR{\mathord{{\eusm S}(\bold R)}}

\def\SRp{\mathord{{\eusm S}^+(\bold R)}}
\def\Kl{\mathord{{\roman K}_\lambda}}
\def\RE{\mathord{\roman{Re}}}
\def\LL{\mathord{{\roman L}^2}}
\def\LKL{\mathord{\LL(\bold R,dx\,\Kl)}}
%
%
\define\thismonth{\ifcase\month 
  \or January\or February\or March\or April\or May\or June%
  \or July\or August\or September\or October\or November%
  \or December\fi}
\let\ul\underline
\def\mint{\int_{\MM} \!d^4x\,}
\def\aeqno#1{\eqno(\roman A.#1)} 
\def\aeqref#1{(A.$#1$)}          
\def\nr{\varkappa^-}
\def\lin#1{\mathopen\langle #1\mathclose\rangle}
\def\ip#1#2{\mathord{\mathopen(#1\mathord{\,\hbox{%
\vrule height 1.5 ex width 0.08 ex depth 0.5 ex}\,}#2\mathclose)}}
\def\sp#1#2{\mathord{\mathopen<#1\mathord{\,\hbox{%
\vrule height 1.4 ex width 0.08 ex depth 0.4 ex}\,}#2\mathclose>}}
\def\os{\mathbin{(\dotplus)}}
\def\IS{\sum_{i=1}^N}
\def\JS{\sum_{j=1}^N}

\def\clot#1{\overline{#1}^\tau}
\def\cc#1{\overline{#1}}
\def\vn{v_{in}}
\def\kint{\int_{-\infty}^{\infty}\!dx\, \Kl(x)}
\def\ISi{\sum_{i=0}^{N-1}}
%

%
%
\topmatter
\title
Mathematical Problems of Gauge Quantum Field Theory:\\
A Survey of the Schwinger Model
\endtitle
\author
by Andreas U. Schmidt\\
\endauthor
\abstract\nofrills{{\bf Abstract.}}
This extended write--up of my talk gives an introductory survey
of mathematical problems 
of the quantization of gauge systems. 
Using the Schwinger model as an exactly tractable but nontrivial example which
exhibits general features of gauge quantum field theory,  I cover the
following subjects: The 
axiomatics of quantum field theory, formulation of quantum field theory 
 in terms of
Wightman functions, reconstruction of the state space, the local formulation 
of gauge theories, indefiniteness of the Wightman functions in general and in 
the special case of the Schwinger model, the state space of the Schwinger 
model,
special features of the model.
\endgraf
New results are contained in the Mathematical Appendix, where I consider
in an abstract setting the Pontrjagin space structure of a 
special class of indefinite inner product spaces --- the so called
{\it quasi--positive} ones. This is motivated by the 
indefinite inner product space structure appearing in the above context and 
generalizes results of Morchio and Strocchi in~\cite{2}, and 
Dubin and Tarski in~\cite{12}.
\endabstract
\endtopmatter
%
%
\topmatter
\title\nofrills{\tenpoint\bf Contents} \endtitle
\endtopmatter
\topmatter
\toc
\eightpoint
\baselineskip0pt
\title 0.~Introduction \endtitle
\title 1.~The Mathematical Setting of Quantum Field Theory \endtitle
\title 2.~The Schwinger Model \endtitle
\title A.~Mathematical Appendix \endtitle
\endtoc
\endtopmatter
\vfill\eject 
\document
%
%
\head 0.~Introduction \endhead
The mathematical problems that gauge quantum field theory raises are so
severe and manifold that not few mathematicians have turned their back
on this part of physics. 
On the other hand the trials to catch the mathematical essence of quantum 
field theory  in
a small set of axioms has led to fundamental insights in the general
structure of such theories.
But it has also shown the difficulties that the quest for a rigoruous 
definition of {\it interacting} theories faces (one should better use
the word {\it impossibility}, see the famous book \cite{SW}).

In this record of my talk I review in a fairly mathematical language a
model theory that gives us some hope --- the Schwinger model, that is,
quantum electrodynamics in two spacetime dimensions.
It is exactly solvable in two senses: One can determine the correlation
functions of the model, as has already been done by Schwinger in \cite{4}.
But from that solution one can also (re--\nobreak) construct the space of 
physical states as a Hilbert space, and this is what will mainly concern
us below.
The mathematically interested reader could take a glance at the
Mathematical Appendix, where I embed part of the problem into the
framework of indefinite inner product spaces and gain some results about
quasi--positive spaces which generalize those of~\cite{2,~12}.

The promising features of the Schwinger model are that shared by other
nontrivial gauge quantum field theories in four dimensions, like for
instance, the confinement of charged particles, nontrivial vacuum structure
and the $\roman U(1)$--problem (see~\cite{1---7}) and one can gain 
appreciable insights in general quantum field theories by investigating 
the model.

The literature on the Schwinger model is vast, and we do not intend here
to give a complete bibliography. Our exposition relies on~\cite{1---3},
but the interested reader should also consult the earlier 
references~\cite{4,~5,~8,~11,~12} and the further results in~\cite{6,~7}.
%
\head 1.~The Mathematical Setting of Quantum Field Theory\endhead
In this section I state the general mathematical stipulations for what follows.
I briefly recall the axiomatic framework of quantum field theory, in its
two formulations. 
%
\subhead The Twofold Axiomatics of Quantum Field Theory\endsubhead
The first is the Hilbert space formulation --- the {\it Wightman axioms}
(see, e.g., \cite{SW}):

{\bf 0. Relativistic Quantum Theory.}~
{\it States} of the theory are unit rays in a separable Hilbert space
$\HH$. There exists a strongly continuous unitary representation $U$ of the
universal covering $\PO$ of the Poincar\'e--group on $\HH$:
$$
\PO\owns\widetilde{\{a,\Lambda\}}\longmapsto U(a,\Lambda)\in{\eusm U}(\HH).
$$
There exists further a {\it unique} translationally invariant state called
the {\it vacuum} represented by an unit vector $\Omega\in\HH$, 
$U(a,1)\Omega=\Omega$, for all $a$ in Minkowski spacetime $\MM\equiv\bold R^4$.
The generators $P^\mu$, $\mu=0,\ldots,3$, of translations (which exist due to
Stones Theorem) shall have
joint spectrum contained in the closed forward lightcone: 
$\spec\{P^\mu\}\subset\forw$ ({\it spectrum condition}).

{\bf I. Fields}~
are a set $\{\varphi_1,\ldots,\varphi_k\}$ of operator valued tempered
distributions. That is, for all test functions $f\in\SM$ are the 
{\it smeared fields}
$$
\varphi_i(f)\equiv\mint \varphi_i(x)f(x):\ \Dom\longrightarrow\Dom
$$
operators on a common domain $\Dom$ of definition.
The vacuum is {\it cyclic} for the polynomial algebra 
$\Falg\equiv\Palg{\varphi_i(f), \varphi_i(f)^\ast|f\in\SM}$ of the
smeared fields: 
$$
\HH=\overline{\Do}\equiv\overline{\Falg \Omega}.
$$

Comment: The above integral notation for the smeared fields is merely to be
understood symbolically. For one can easily show (see, e.g., \cite{1}, sect. 1.2)
that the only way to define fields as {\it functions}, which is compatible with
axioms 0 --- III, is to take trivial fields $\varphi_i(f)=\lambda\boldkey 1$.

{\bf II. Poincar\'e--Covariance.}~
The fields transform {\it covariantly} under the adjoint action of $U$:
$$
U(a,\Lambda)\varphi_i(x)U(a,\Lambda)^{-1}=
{S^k}_j(\widetilde{\Lambda}^{-1})\varphi_k(\Lambda x+a).
$$
Here $S$ is a finite--dimensional representation of $\roman{SL}(2,\bold C)$,
the universal covering of the proper orthochronuous Lorentz group $\POLG$.
The above equation shall hold in the strong sense on $\Dom$, when
both sides are smeared with $f\in\SM$.

{\bf III. Locality/Causality.}~
If $\supp f$ is spacelike to $\supp g$, $f,g\in\SM$ (i.e.,
$(x-y)^2<0,\ \forall x\in\supp f,\ y\in\supp g$),
then the smeared fields $\varphi_i(f)$ and $\varphi_j(g)$ either
commute or anticommute: $\komm{\varphi_i(f)}{\varphi_j(g)}_\pm=0$,
representing the {\it Bose--Fermi--alternative}.

The second formulation is that in terms of vacuum expectaion values or
{\it Wightman functions}. First, one point about notation.
We introduce the space $\SB$ of {\it test function sequences}
$$
\SB\equiv\left\{ \ul f=(f_0,f_1,\ldots)|f_0\in\bold C,\,f_i\in\SMi,\,
\#\{i,f_i\neq0\}<\infty\right\},
$$
which will play an important role in our discussion of the reconstruction
theorem below. Now to the second set of axioms for relativistic
quantum field theory.

{\bf A. Relativistic $\boldkey n$--Point--Functions.}
There exist tempered distributions, also called {\it correlation functions},
$\WW_n\equiv\WW(x_1,\ldots,x_n)\in\SMn^\prime$
for all $\mathord{n\in\bold N}$. The $\WW_n$ are 
{\it translationally invariant},
i.e., there exist distributions $\Wf_{n}$ such that
$$
\WW(x_1,\ldots,x_n)=\Wf(\xi_1,\ldots,\xi_{n-1}),\quad\xi_i=x_{i+1}-x_i.
$$
Further, the $n$--point-functions  shall be {\it Lorentz--invariant}:
$\!\WW(\Lambda x_1,\ldots,\Lambda x_n)\!=$ $\WW(x_1,\ldots,x_n)$, 
$\forall\Lambda\in\POLG$. A {\it spectrum condition} is demanded, i.e.,
the Fourier--transforms $\widetilde{\Wf}_n$ shall have their support
in the $n-1$--fold product of the forward lightcone:
$\supp\widetilde{\Wf}_n\subset\forw^{n-1}$. Another important condition 
imposed is that of
{\it hermiticity}. For $f_n\in\SMn$ define $f_n^\ast(x_1,\ldots,x_n)\equiv$
$\cc{f_n}(x_n,\ldots,x_1)$. Then
$$
  \WW_n(f_n)=\cc{\WW_n(f_n^\ast)}.
$$

{\bf B. Positity} 
means that, for $\ul f, \ul g\in\SB$, the {\it sesquilinear form} given by
$$
  \sp{\ul f}{\ul g}\equiv\sum_{m,n}\WW_{m+n}(f_n^\ast\otimes g_m)
\eqno(1.1)
$$
shall be positive, i.e., $\sp{\ul f}{\ul f}\geq 0$.
(We have set 
$f_n\otimes g_m(x_1,\ldots,x_{m+n})\equiv$
$f_n(x_1,\ldots,x_n)g_m(x_{n+1},\ldots,x_{m+n})$).

{\bf C. Cluster Property.} 
For every spacelike vector $a\in\MM$ and 
$\bold R \owns \lambda\rightarrow \infty$  the condition
$$
\lim_{\lambda\rightarrow\infty}
\WW(x_1,\ldots\!,x_j,x_{j+1}+\lambda a,\ldots\!,x_n+\lambda a)-
\WW(x_1,\ldots\!,x_j)\WW(x_{j+1},\ldots\!,x_n)\!=\!0
$$
shall hold in the sense of distributions.

Comment: This is a less intuitive axiom from the mathematical standpoint.
It means roughly that the correlation of clusters of fields, which become
infinitely spacelike separated, factorizes (i.e., a decorrelation takes place).
This is tied to the independence of events in the two clusters in that limit
and motivated by results from (Haag--Ruelle) scattering theory (see, e.g.,
\cite{BLT}). It is remarkable that the cluster property serves also to
ensure the uniqueness of the vacuum state in the reconstruction theorem 
discussed below.

{\bf D. Local Commutativity.} Whenever $(x_j-x_{j+1})^2<0$,
$$\WW(x_1,\ldots,x_j,x_{j+1},\ldots,x_n)=
\WW(x_1,\ldots,x_{j+1},x_j,\ldots,x_n)$$
follows.
%
\subhead Equivalence of the Two Formulations\endsubhead
It is relatively clear that we can gain $n$--point functions by the simple 
definition of vacuum expectation values
$$
\WW_n(x_1,\ldots,x_n)\equiv\sp{\Omega}{\phi(x_1)\ldots\phi(x_n)\Omega}.
$$
(For simplicity we consider only one single hermitean scalar field in all
the following.
Otherwise, one would have to take more than one sort of Wightman functions
into account, i.e., one would have to go over to tensor products of
distribution
spaces). With the above definition, the properties A --- D follow rather 
directly from 0 --- III (see, e.g., \cite{SW}). The idea how to reconstruct
the theory, in terms of a Hilbert space and fields, from the Wightman 
functions, is also simple but mathematically a bit more involved. Roughly, one
uses the space $\SB$ as the raw material for building the Hilbert space $\HH$.
%
\subhead Sketch of the Reconstruction Theorem\endsubhead
On $\SB$ one immediately has a linear structure given by
$(\alpha\ul f+\beta\ul g)_i\equiv\alpha(\ul f)_i+\beta(\ul g)_i$, for
$\alpha,\beta\in\bold C$. A linear representation of the Poincar\'e--group
is induced by setting
$$
(U(a,\Lambda)\ul f)_i\equiv(\ul f)_i(\Lambda^{-1}(x-a)).
\eqno(1.2)
$$
The positivity condition B gives the obvious idea, to take 
the form $\sp . .$ as a candidate for a scalar product on $\SB$.
What we shall call fields will then act simply as a sort of
``creation operators''
on $\SB$: That is, for all $h\in\SM$ the action of the map 
$\varphi(h):\SB\rightarrow\SB$ is defined by
$$
\varphi(h)\ul f\equiv (0,h f_0,h\otimes f_1,\ldots).
\eqno(1.3)
$$
The condition of translational invariance formulated in 0 gives a simple
choice for a vacuum state, namely
$\Omega\equiv(c,0,0,\ldots)$, $c\in\bold C$.
The main technical hurdle to take is the fibering of $\SB$ into equivalence 
classes and subsequent Hilbert space completion. One has to quotient out the
isotropic part $\SB^\bot$ (see the appendix for its definition) 
to render the scalar product positive definite, and finally arrives in 
Hilbert space:
$$
\HH\equiv\cc{\SB\mathbin{\Bigr/}\SB^\bot}.
$$

Some comments about the interdependence of the axioms: First, the definition
($1.1$) together with the invariance properties of the $\WW_n$ makes
the representation ($1.2$) unitary and the fields defined by ($1.3$) will
transform under the adjoint action of $U$ (for the case of a single scalar 
field), thus they fulfill II. Locality  follows directly from D. The only 
remarkable thing to note is, that the cluster property C is necessarily needed
to show that the vacuum is the only translationally invariant state.
To see a simple physical argument for that the reader might take a look
at \cite{1}, pp. 15. But see also the complete proof of the reconstruction 
theorem in \cite{SW}, chapter III.
%
%
\head 2.~The Schwinger Model \endhead
%
%
\subhead Definition of the Model\endsubhead
Now we will introduce the model we are about to consider, in the symbolic 
notation common to physicists.

The {\it Schwinger model}, also called $\roman{QED}_2$, is a 
Quantum Field Theory on two--dimensional Minkowski--spacetime, i.e.,
$\bold R^2$, $x=(x^0,x^1)$, with diagonal metric $g^{00}=-g^{11}=1$. 
Let $\varepsilon$ be
the antisymmetric rank--$2$--tensor, $\varepsilon^{10}=-\varepsilon^{01}=1$,
and define the two--dimensional gamma matrices by
$$
\gamma^0\equiv
\left(\matrix 0 & 1 \cr 1 & 0 \cr \endmatrix\right),\quad
\gamma^1\equiv
\left(\matrix 0 & 1 \cr -1 & 0 \cr \endmatrix\right),\quad
\gamma^5\equiv\gamma^0\gamma^1.
$$
The ingredients of the theory are two fields, as usual in quantum 
electrodynamics: First, a {\it fermion field}
$$
\psi\equiv\left(\matrix \psi_1 \cr \psi_2 \cr\endmatrix\right),\quad
\cc{\psi}\equiv\psi^\ast\gamma^0.
$$
Second, a {\it gauge potential} $A_\mu$ --- a two--dimensional vector field
with affiliated {\it field strength}
$$
F_{\mu\nu}\equiv\partial_\mu A_\nu-\partial_\nu A_\mu.
$$
The {\it equations of motion} are the {\it Dirac equation}
$$
i\gamma^\mu\partial_\mu\psi+g\gamma^\mu[A_\mu\psi]_{\roman{ren}}=0,
\eqno(2.1)
$$
where $g$ is the {\it coupling constant} (replacing the electric charge
unit in usual QED), and the {\it local Gau\ss\ law}
$$
\partial^\nu F_{\nu\mu}=-g j_\mu,
\eqno(2.2)
$$
where $j_\mu$ is the {\it ``electric'' current} to be defined below.

The product of fields $A_\mu\psi$ at a point in spacetime is, at first,
an ill--defined object as well as the current operator $j_\mu$.
Namely, when one imposes the canonical commutation relations on the 
fields $A_\mu$, $\psi$ and their conjugated momenta, one immediately 
encounters singularities if one tries to define the operator products
in a na\"\i ve way.  
To give sense to them as operator--valued distributions one has to adopt
a regularization procedure. It reads
$$
j_\mu(x)\mathord\equiv\!
\lim_{\scriptstyle{\varepsilon\rightarrow0}\atop\scriptstyle{\varepsilon^2<0}}
\left\{
  \cc{\psi}(x\mathord+\varepsilon)\gamma_\mu\psi(x)\mathord-
  \sp{\Omega}{\cc{\psi}(x\mathord+\varepsilon)\gamma_\mu\psi(x)\Omega}
\right\}
\left[
  1\mathord-ig\varepsilon^\nu A_\nu(x)
\right]
\eqno(2.3)
$$
for the current. The last term makes the procedure of subtracting the 
singularity, which is given by the vacuum expectation value,  gauge invariant. 
And for the renormalized operator product one sets
$$
[A_\mu(x)\psi(x)]_{\roman{ren}}\equiv
\lim_{\scriptstyle{\varepsilon\rightarrow0}\atop\scriptstyle{\varepsilon^2<0}}
  {1 \over 2}
\bigl\{
A_\mu(x+\varepsilon)\psi(x)+\psi(x)A_\mu(x-\varepsilon)
\bigr\}.
\eqno(2.4)
$$
 
At this point a comment about the mathematical status of the above notions
is at hand. They are ``hypothetical,'' in the sense that there is a priori
no Hilbert space of states in which the above expressions could be defined
as weak operator limits. Nevertheless, by the canonical quantization procedure 
and by the Lagrange--equation of the model, one has enough data to determine
the singularities exact enough to subtract (as in ($2.3$)) or regularize
them away (as in ($2.4$)).

To complete this ``bootstrap approach'' the central task is the construction of
the state space. Before addressing this issue in the context of the Schwinger
model, I will first mention a general problem, which every gauge 
quantum field theory raises. In this I will assume that the entities
``vacuum'' and ``Hilbert space,'' in the sense of the Wightman axioms, exist.
%
\subhead Local Formulation of Gauge Theories \endsubhead
The main question in gauge quantum field theory is: Can one construct 
{\it charged states}, and is it possible to give sense to {\it charge 
operators} as observables? Charged states are classically characterized by
the Gau\ss\ law, i.e., the conservation of the Noether current corresponding
to the gauge symmetry. The first best guess for a quantum charge operator
is to smear out the zero--component of the current (in the meantime
I am talking about general QFT in four dimensions):
$$
Q_R\equiv
\mint j_0(\boldkey x,t)f_r(\boldkey x)\alpha(t),
\eqno(2.5)
$$
where $f_R(\boldkey x)\equiv f(|\boldkey x|/R)$, with $f$ a test function
obeying $f(\boldkey x)=1$ for $|\boldkey x|<1$, $f(\boldkey x)=0$ for
$|\boldkey x|>1+\varepsilon$ and $\alpha\in\eusm D(\bold R)$, $\int\!dt\,
\alpha(t)=1$. In a quantum theory this charge should fulfill the commutation 
relations of a generator of gauge transformations
$$
\lim_{R\rightarrow\infty}[Q_R\mathord,A]=qA
$$
with every charged field $A$ carrying total charge $q$, which is a scalar
in the abelian case, as we assume here. 

The problem one has to face now is
that a theory in which the local Gau\ss\ law holds will  not contain
any {\it local} charged fields. Namely one has 
$$
[Q_R,A]=\mint f_R(\boldkey x)\alpha(t)[\partial^iF_{i0}\mathord,A]=
\mint\partial^if_R(\boldkey x)\alpha(t)[F_{i0}\mathord,A]
$$
by antisymmetry of $F$ and ($2.2$). Now the function $\partial^if_R\alpha$ has
support in the shell $R<|\boldkey x|<R(1+\varepsilon)$ and compact support
in time direction. If $A$ is a local observable in the sense of axiom III, then
its own support will be bounded and become spacelike to the support
of $\partial^if_R\alpha$ for large enough $R$. It follows that
$\lim_{R\rightarrow\infty}[Q_R\mathord,A]=0$ and $A$ has zero charge.

There are two ways to overcome this: Either one could use non--local fields
or one could retain locality and modify Gau\ss' law. The second approach 
is what we call the {\it local formulation of gauge QFT's}: We insist on 
locality of all fields on the expense of introducing an unphysical 
{\it longitudinal field} $j_\mu^{\roman L}$ (a term coming from QED), 
which adds to the current in ($2.2$). The result is the {\it modified 
Gau\ss\ law}
$$
\partial^\nu F_{\nu\mu}=-g j_\mu+gj_\mu^{\roman L}.
\eqno(2.6)
$$
Clearly, the effect of the additional term is to give a nontrivial contribution
for the quantum charge. But, as $j_\mu^{\roman L}$ is an artifact of 
quantization, it should have no observable content by itself. That is, one
has to impose the condition
$$
\sp{\Phi}{j_\mu^{\roman L}\Psi}=0,\quad \forall\Phi,\Psi\in\HPH,
\eqno(2.7)
$$
on the vacuum expectation values of states in the {\it physical} Hilbert space
$\HPH$, which has yet to be defined.
Equation ($2.6$) together with condition ($2.7$) is called {\it weak Gau\ss\ 
law}.
The subspace $\HPH$ is singled out from the total space $\HH$ (which is 
created by acting on the vacuum by {\it all} fields, even the unphysical)
by the so--called {\it subsidiary} or {\it BRST--condition}
$$
Q_{\roman{BRST}}\equiv(j_\mu^{\roman L})^-\Psi=0,\quad \forall\Psi\in\HPH.
\eqno(2.8)
$$
Here $(.)^-$ means taking the negative--energy--, i.e., annihilator--part 
of a given field. This suffices to ensure ($2.7$), as one easily sees
using hermiticity of $j_\mu^{\roman L}$.

There is, however, a price to be paid for the local formulation: The Wightman 
functions will, under very general conditions, no longer satisfy the positivity
axiom B. For it holds
\prop{}
A local formulation of a gauge theory satisfying all Wightman axioms 0 --- III
contains no charged states in $\HPH$.
\endprop
{\sproof{}
From $\Omega\in\HPH$ and the fact that $j_\mu$ and $F_{\mu\nu}$ are 
observable fields, it follows that 
$$
j_\mu^{\roman L}(f)\Omega=
\mathord{\left(j_\mu+{1 \over g}\partial^\nu F_{\nu\mu}\right)}(f)
\Omega\in\HPH.
$$
Then the weak Gau\ss\ law implies 
$\sp{j_\mu^{\roman L}(f)\Omega}{j_\mu^{\roman L}(f)\Omega}=0$, 
$\forall f\in\SM$. Since the vacuum is not only cyclic but also a separating
vector for the field algebra (an important consequence of the Wightman axioms
called the {\it Reeh--Schlieder--Theorem}, cf. \cite{SW}), it follows
that $j_\mu^{\roman L}(f)\Omega=0$, {\it provided positivity holds}.
If now $j_\mu^{\roman L}(f)$ is localized in a bounded region 
$\eusm O\subset\MM$ and $P_1$, $P_2$ are polynomials in fields localized
in regions $\eusm O_1$, $\eusm O_2$ respectively, both spacelike
separated from $\eusm O$, then 
$\sp{P_1\Omega}{j_\mu^{\roman L}(f)P_2\Omega}=
\sp{P_1\Omega}{P_2j_\mu^{\roman L}(f)\Omega}=0.$
Since the states generated by polynomials in the local fields form a
dense subset in $\HH$ (Axiom 0), we conclude that $j_\mu^{\roman L}(f)=0$
as an operator in $\HH$. This means nothing but that the Gau\ss\ law holds
in its original form, which, as we have seen above, implies that $\HPH$
contains no charged states.
\endsproof}
It should be clear from the above proof that indeed the properties in conflict
are locality and positivity. This means that the space generated by the 
polynomial algebra $\eusm P$ of the local fields from the vacuum must be
an indefinite inner product space $\VV\equiv\eusm P\Omega$, and the physical 
Hilbert space should be the positive part of this space determined by 
condition ($2.8$). 

Now that we have saved locality of our fields and possibly have gained by
the subsidiary condition a physical space $\HPH$, the next question to
ask (which is mostly neglected in literature, see the discussion in \cite{1},
pp. 91) runs: Is $\VV$ modulo ($2.8$) large enough to contain charged states? 
Unfortunately the answer is negative, at least for abelian gauge theories
(and there seems to be no reason why this should become better in non--abelian
theories), as the following result shows.
\prop{\cite{1, Prop. 6.4}}
In a local formulation of QED all the physical local states have zero charge.
\endprop
The situation so far is sketched below:
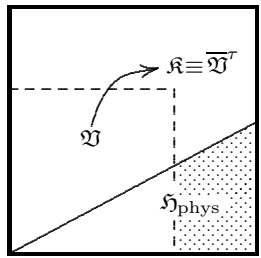
\input prepictex
\input pictex 
\input postpictex
$$
\beginpicture 
\eightpoint
\setcoordinatesystem units <1mm,1mm>
\setplotarea x from -5 to 35, y from 0 to 30
\linethickness1pt
\putrectangle corners at 0 0 and 30 30
\setdashes <1mm>
\plot 0 20 20 20 20 8 /
\plot 20 4 20 0 /
\setsolid
\linethickness0.8pt
\plot 0 0 30 16 /
\put {$\VV$} [tr] at 11 15 
\put {$\KK\!\equiv\!\clot{\VV}$} [bl] at 19 22
\put {$\HPH$} [cc] at 22 6
\arrow <2mm> [0.25,0.75] from 15 22 to 18 22.5
\setquadratic
\plot 15 22 12 20 10 15.5 /
\setlinear
\setshadegrid span <1.6pt>
\vshade 20 7 10.67 30 7 16 /
\vshade 20 0 4 30 0 4 /
\vshade 23 4 1.67 30 4 10.67 /
\endpicture
$$
We have an indefinite inner product space $\VV$
(we assume $\VV$ to be nondegenerate, otherwise one would
have to quotient away its isotropic part)
 containing physical as well
as unphysical states, but no charged states since $\VV$ is generated from
the vacuum by polynomials in {\it local} fields. To find charged states as
suitable limits of states in $\VV$, one has to look for a completion of $\VV$
in an appropriate topology $\tau$. After that, one should impose
the subsidiary condition to get the physical Hilbert space $\HPH$.
The charged and presumably nonlocal states are then contained in the
shaded area of the above figure.

In our quest of a topology the framework of majorant Hilbert topologies and
Krein spaces, as developed in the first part of the appendix, comes in place
(consult this section for definitions).
Let us first replace the by now unsufficient positivity condition B by
the weaker assumption

{\bf B${}^\prime$. Hilbert space structure condition.}
Assume that there exist Hilbert seminorms $p_n$ on $\SMn$ such that
$$
\bigl|\WW_{n+m}(f_n^\ast\otimes g_m)\bigr|\leq p_n(f_n)p_m(g_m)
\eqno(2.9)
$$
for all $f_n\in\SMn$, $g_m\in\SMm$.

Taking a look at lemma A.7 in the appendix we see that, in a theory where
B${}^\prime$ holds, $\VV$ will possess a majorant Hilbert topology $\tau$
induced by a single majorant Hilbert norm $p$.
Then A.9 --- A.12 tell us that there exists a completion of 
$\VV$ w.r.t. a minimal majorant Hilbert topology $\tau_\ast$ to a
Krein space $\KK$, i.e., a maximal Hilbert space structure $(\KK,G)$,
with metric operator $G$, for $\VV$. This is the best we can do with
indefinite inner product spaces.
Let us now see how B${}^\prime$ works in the concrete example.
%
\subhead The State Space of the Schwinger Model\endsubhead
Lowenstein and Swieca have shown in \cite{5} that the solution
of $\roman{QED}_2$ in terms of Wightman functions in the original paper
\cite{4} can be rewritten formally in terms of the so--called {\it building
block fields}. These are three free fields $\eta$, $\Sigma$ and $\psi_0$.
$\psi_0(x)$ is a massless free Dirac fermion field. $\Sigma(x)$ is a massive
scalar field of mass $M=g/\sqrt{\pi}$ obeying the equation of motion
$$
\square\Sigma(x)=M^2\Sigma(x).
$$
Finally $\eta$ is a massless scalar field, on which we will concentrate our
attention below. First, let us write down the formal solution to ($2.1$) and
($2.6$) in terms of these fields. It reads
$$
\eqalign{
\psi(x)&=\mathopen\boldkey{:}e^{i\sqrt{\pi}\gamma^5(\eta+\Sigma)}
\mathclose\boldkey{:}(x)\psi_0(x),\cr
A_\mu(x)&=-{\sqrt{\pi}\over g}\varepsilon_{\mu\nu}
\bigl\{
  \partial^\nu\eta(x)+\partial^\nu\Sigma(x)
\bigr\}.
\cr}
\eqno(2.10)
$$
Here the Wick--ordered exponential in the expression for $\psi(x)$ is defined
as usual by its formal power series (see, e.g., \cite{BLT}). Using the 
equations of motion for $\eta$ and $\Sigma$, and inserting ($2.10$) into
the regularization prescription ($2.3$) for the current, one gets explicit 
expressions for the physical and the longitudinal current in ($2.6$)
(see \cite{5} for details):
$$
\eqalign{
j_\mu(x)&=j_\mu^{\roman{free}}+{g\over\pi}A_\mu(x),\quad\roman{where}\ 
j_\mu^{\roman{free}}(x)\equiv
\mathopen\boldkey{:}\psi_0\gamma_\mu\psi_0\mathclose\boldkey{:}(x),\quad
\roman{and}\cr
j_\mu^{\roman L}(x)&=j_\mu^{\roman{free}}(x)-{1\over\sqrt{\pi}}
\varepsilon_{\mu\nu}\partial^\nu\eta(x).
\cr}
$$

Let us reconsider the nonpositivity problem above in the example.
As the two--point functions of the free fields in question are well known,
one can, at first, conclude that nonpositivity can only come from the 
two--point function of the massless scalar field $\eta$. For   
it has the special form in two dimensions  
(see \cite{1} and \cite{8} for definitions and notations):
$$
\ip{\eta(x)\Omega}{\eta(y)\Omega}=-D^+(x-y),\ \roman{where}\ 
D^+(x)\equiv\lim_{\varepsilon\searrow0}{1\over 4\pi}
\ln\left(x^2+i\varepsilon x^0\right).\eqno(2.11)
$$
The limit has been taken to define the expression as a
distribution on $\SRR$ and we have used the notation $\ip . .$ for the 
indefinite form. 
The indefiniteness can be made explicit using the momentum--space
representation of $D^+(x)$.
Going over to {\it lightcone coordinates} $p_\pm\equiv p_0\pm p_1$,
one finds that the Fourier--transform
$\widetilde{D^+}(p)$ is given by
$$
\widetilde{D^+}(p)=
\left[
\left({1\over p_+}\right)_+\!\delta(p_-)+
\left({1\over p_-}\right)_+\!\delta(p_+)
\right] 
\Theta(p_0)
\eqno(2.12)
$$
(see \cite{8}). ($2.12$) is --- up to a constant which we suppressed --- 
the most general Lorentz invariant Distribution
satisfying $p^2\widetilde{D^+}(p)=0$, and concentrated on the
plus lightcone $C^+$, the surface on which $p^2=0$, $p_0\geq0$. 
In this expression we used the canonically regularized distribution
(cf. \cite{GS}, \S I.3)
$$
\left({1\over p}\right)_+\!(\widetilde{f})\equiv
\int_0^\infty\!dp\,{\widetilde{f}(p)-\widetilde{f}(0) \over p}.
$$
The necessity of a regularization at $p=0$ makes it clear that
the two--point function will violate the positivity postulate B 
for general $f\in\SRR$,
with $\widetilde{f}(0)\neq0$.

Now we can again apply the results of the appendix ---
especially of the last section --- to reconstruct the state space of
the theory from its Wightman functions, as sketched in section 1. This will
now actually be a Krein space.
We begin with the subspace $\SRR$ of $\ul{\eusm S}(\bold R^2)$ corresponding
to the part of the inner product coming from the two--point function.
On this, the indefinite inner product \aeqref{5} has the special form 
$$
\ip{f}{g}_{(1)}\equiv
-D^+\left(\cc f (x) g(y)\right)=
-\widetilde{D^+}\left(\cc{\widetilde{f}}(-p) \widetilde{g}(p)\right).
$$
By ($2.12$), the singular integral kernel $\Kl$ of \aeqref{4} is 
$1/p_\pm$ acting on two copies of the real half--axis $[0,\infty)$.
One finds a suitable test function $\chi_0\in\SRR$ with Fourier--transform
that is unity at the origin, $\widetilde{\chi_0}(0)=1$, and
$\ip{\chi_0}{\chi_0}_{(1)}=0$.
Then the linear decomposition of $\SRR$ as in \aeqref{6},
\aeqref{7} takes place. The Fourier--transform of every $f\in\SRR$ decomposes
into
$$
\widetilde{f}=\widetilde{f^+}+\chi_0\widetilde{f}(0),
$$
with a function $f^+\in\SRR$, satisfying $\widetilde{f^+}(0)=0$.
We find a majorant Hilbert seminorm $p_\eta$
and a positive inner product ${\sp . .}_{(1)}$ on $\SRR$ as in \aeqref{8}.
It is explicitly given by
$$
{\sp f g}_{(1)}\equiv-\ip{f^+}{g^+}_{(1)}+
{\ip{f}{\chi_0}}_{(1)}{\ip{\chi_0}{g}}_{(1)}+
\cc{\widetilde{f}(0)}\widetilde{g}(0).
$$
The minus sign in the first term is compensating the one in the expression
($2.11$) for the two--point function (we are actually dealing here
with a quasi--negative instead of a quasi--positive space, but this
poses no problem for the general procedure).
Finally, the completion of $\SRR$ w.r.t. the topology $\tau(\eta)$
induced by $p_\eta$ will be 
$$
\KK^{(1)}_\eta\equiv\cc{\SRR\mathbin{\Bigr/}\SRR^\bot}^{\widehat{\tau(\eta)}}=
\roman L^2\left(C^+,{dp\over |p|}\right)
\oplus\lin{v_0}\oplus\lin{\chi_0},
$$
like in \aeqref{9}, 
where $\widehat{\tau(\eta)}$ is the induced topology on the 
quotient. 
This is what we could call the ``one--particle space'' 
of $\eta$. 
It is a Pontrjagin space with rank of positivity equal to one.
Since the $n$--point functions of a free scalar field
decompose tensorially into sums of products of the two--point function,
one can render the total {\it Fock space} of this field as an orthogonal sum
of tensor products
$$
\KK_\eta=\bigoplus_n\left(\KK^{(1)}_\eta\right)^{\otimes n}.
$$

Further analysis of the state--space structure yields: $p_\eta$, together with 
the positive seminorms given by the two--point functions of $\Sigma$ and
$\psi_0$, induces a minimal majorant Hilbert topology $\tau$ on the space 
$\eusm F\Omega$, generated from the Fock vacuum by the action of the polynomial
algebra $\eusm F=\Palg{A_\mu,\psi}$ of the physical fields. Setting
$\KK\equiv\clot{\eusm F\Omega}$, one finds (see \cite{1, 3})
$$
\KK=\HH_\Sigma\otimes\KK_{\eta,\psi_0}.
$$
Here $\HH_\Sigma$ is the Fock space of $\Sigma$, and 
$\KK_{\eta,\psi_0}$ is $\clot{\eusm F_{\eta,\psi_0}\Omega}$,
where 
$$
\eusm F_{\eta,\psi_0}\equiv\Palg{j_\mu^{\roman{free}},\,j_\mu^{\roman{L}},\,
\mathopen\boldkey{:}e^{i\sqrt{\pi}\gamma^5\eta}
\mathclose\boldkey{:}\psi_0}
$$
is the field algebra generated by all physical fields not containing
$\Sigma$. The last step is to impose the subsidiary condition to get the 
physical Hilbert space. Denote by
$\widehat{\KK}\equiv\left\{\Psi\in\KK|\,(j_\mu^{\roman{L}})^-\Psi=0\right\}$
the subspace of solutions to this condition. Then finally,
$$
\HPH=\cc{\widehat{\KK}\mathbin{\Bigr/}\widehat{\KK}^\bot}^{\widehat{\tau}}.
$$
%
\subhead Physical Features of the Model\endsubhead
Coming to the end, let me say a few words about the interesting features
of the Schwinger model, which can be rigoruously derived, now that we have
identified its physical state space. These features are interesting because 
they are shared by nontrivial gauge quantum field theories in four dimensions
like QCD.

The first to mention is the {\it confinement of charged particles}. That is,
if we define the ``electric charge'' as in ($2.5$) by 
$$
Q_R^{\roman{el}}\equiv
\int_{\bold R^2}\!d^2x\,j_0(x^0,x^1)f_R(x^1)\alpha(x^0),
$$
then
$$
\mathop{\roman{s\mathord-lim}}\limits_{R\rightarrow\infty}
\exp\left({i\lambda Q_R^{\roman{el}}}\right)
=\boldkey 1_{\HPH},\quad\forall\lambda\in\bold C
$$
($\mathop{\roman{s\mathord-lim}}$ denotes the limit in the strong operator
sense).
That means, the electric charge vanishes on the physical space --- charges
are ``invisible.''

Much more would be left to tell about, e.g., the {\it nontrivial vacuum 
structure}, the {\it energy gap}  and the {\it $\roman U(1)$--problem},
but lack of space urges me to stop here. The physically interested reader
might consider to take a look at \cite{1, 3, 6 and 7} for discussions of these 
issues.
%
\head A. Mathematical Appendix \endhead
%
\subhead Generalities About Indefinite Inner Product Spaces \endsubhead
In this subsection we recall some facts about indefinite inner product, 
Krein and Pontrjagin spaces needed in the subsequent sections and the
main text. This is 
just to be self--contained. For an extensive discussion of the subject 
see \cite{BOG}.

First some notations: 
Let $\VV$ be a vector space equipped with an indefinite inner product
$\ip{.}{.}$
(antilinear in the first, linear in the second argument). 
The linear span of a subset $\SPACE{A}$ of vectors in $\VV$
is denoted by $\lin{\SPACE{A}}$. The {\it linear sum} of subspaces 
$\VV_1,\ldots,\VV_n$ of $\VV$ is given by $\lin{\VV_1\cup\cdots\cup\VV_n}$
and denoted by $\VV_1+\cdots+\VV_n$. If the spaces $\VV_1,\ldots,\VV_n$
are linearly independent, their linear sum is termed {\it direct sum}
and denoted by $\VV_1\dotplus\cdots\dotplus\VV_n$. {\it Orthogonality} w.r.t.
$\ip{.}{.}$ is defined, and denoted by the binary relation $\bot$ as usual
(but clearly does not have the same strong consequences as in definite
inner product spaces). If the $\VV_1,\ldots,\VV_n$ are mutually orthogonal,
their {\it orthogonal direct sum} is denoted by $\VV_1\os\cdots\os\VV_n$,
whereas the symbol $\oplus$ is reserved for orthogonal sums w.r.t. a 
positive definite inner product, which we will casually denote with
$\sp{.}{.}$. By {\it positive definite} we mean as usual
$\sp x x \geq 0$, $\forall x \neq 0$, and $\sp x x =0 \Rightarrow x=0$.
A subspace $\SPACE{A}$ of $\VV$ is called {\it positve}, 
{\it negative}, or {\it neutral}, respectively, if one of the possibilities
$\ip x x > 0$, $\ip x x <0$ or $\ip x x =0$ holds for all $x\in\SPACE A$,
$x\neq 0$.
One sets
$$
\Vpp\equiv\{x\in\VV|\,\ip x x > 0\ \roman{or}\ x=0\},
$$
and calls this subset the {\it positive part} of $\VV$. The {\it negative}
and {\it neutral} parts $\Vmm$ and $\VV^0$ are defined alike.
A subspace $\SPACE A$ of $\VV$ is called {\it degenerate}, if its
{\it isotropic part} $\SPACE A \cap \SPACE{A}^\bot$ does not only consist
of the zero vector. In the 
following we will deal merely with {\it non--degenerate} spaces, i.e.,
spaces with $\VV^\bot=\{0\}$.
\lem{A.1 \cite{BOG, Lemma I.2.1}}
Every indefinite inner product space contains at least one nonzero neutral
vector.
\endlem
\lem{A.2 \cite{BOG, Cor. I.4.6}}
If two vectors in an inner product space satisfy $\ip x y \neq 0$, 
$\ip x x =0$, then the subspace $\lin{x,y}$ is indefinite.
\endlem
A non--degenerate inner product space $\VV$ is said to be {\it decomposable} 
if it admits a {\it fundamental decomposition}
$$
\VV=\VV^\bot\os\Vp\os\Vm,\quad \roman{with}\ \Vp\subset\Vpp,\ \Vm\subset\Vmm.
$$
For non--degenerate spaces the isotropic part of the decomposition vanishes.
The special species of decomposable inner product spaces that we will consider
in the next section is defined as follows: 
\defi{A.3}
$\VV$ is called a 
{\it quasi--positive} ({\it quasi--negative}) inner pro\-duct space, if it does
not contain any negative definite (positive definite) subspace of infinite
dimension.
\enddefi
\lem{A.4 \cite{BOG, Thm. I.11.7}}
Every quasi--positive (quasi--negative) inner product space is decomposable.
\endlem
The dimension of a maximal negative definite subspace
$\Vm\subset\Vmm$ of a non--degenerate, quasi--positive space is called the 
{\it rank of negativity} of $\VV$. It is an unique positive cardinal 
(\cite{BOG, Cor. II.10.4}) denoted by $\nr(\VV)$. The {\it rank of
positivity} $\varkappa^+(\VV)$ is defined in analogy to that. We set
$\varkappa\equiv\min\{\nr(\VV),\varkappa^+(\VV)\}$ and call this number
the {\it rank of indefiniteness} of $\VV$. 

Now some less trivial things about the topology of indefinite inner product 
spaces: A locally convex topology $\tau$ on $\VV$ defined by a single 
seminorm $p$, which is then actually a norm, is called {\it normed}.
If $\VV$ is $\tau$--complete, we say that $\tau$ is a 
{\it Banach topology}. If
$\tau$ can be defined by a {\it quadratic norm} $p(x)=\sp{x}{x}^{1/2}$,
where $\sp{.}{.}$ is a positive inner product on $\VV$, then $\tau$ is called
a {\it quadratic normed topology}. Again, if $\VV$ is $\tau$--complete,  
then $\tau$ is termed
{\it Hilbert topology}. A normed topology $\tau_1$ is {\it stronger}
than another $\tau_2$, written $\tau_1\geq\tau_2$, iff every $\tau_2$--open
set is also a $\tau_1$--open set. This is the case, iff the relation 
$p_1(x)\geq\alpha p_2(x)$ holds, with $\alpha>0$ for all $x\in\VV$. Two
norms that define the same topology are called {\it equivalent}.

A locally convex topology $\tau$ on $\VV$ is called a {\it partial majorant}
of the inner product, iff $\ip . .$ is separately $\tau$--continuous. 
The {\it weak topology} on $\VV$ is the topology defined by the family of 
seminorms
$$
p_y(x)\equiv|\ip{y}{x}|,\quad\forall x\in\VV.
$$
\lem{A.5 \cite{BOG, Thm. II.2.1}}
The weak topology is the weakest partial majorant on $\VV$.
If a locally convex topology on $\VV$ is stronger than the weak topology,
then it is a partial majorant.
\endlem
Below we will encounter the stronger concept of a {\it majorant
topology}:
\defi{A.6}
A locally convex topology $\tau$ on $\VV$ is called {\it majorant topology}, if
the inner product $\ip{.}{..}$ is jointly $\tau$--continuous. 
\enddefi
The following properties of majorant topologies were used in section 2,
equation ($2.9$), to
modify the Wightman axioms for local gauge theories.
\lem{A.7 \cite{BOG, Lemma IV.1.1 \& 1.2}}
\roster
  \item "i)" To every majorant there exists a weaker majorant defined by
    a single seminorm.
  \item "ii)" For a locally convex topology defined by a single seminorm
    $p$ to be a majorant it is sufficient that $p$ dominates the inner
    square:
    $$
    |\ip x x | \leq \alpha p(x)^2,\quad\alpha>0,\ \forall x\in\VV.
    $$
\endroster 
\endlem
Majorant topologies --- especially majorant Hilbert topologies --- have
many advantages over partial majorants. Before we describe them, let us see
why one would not like to use the weak topology on  general
indefinite inner product spaces:
\lem{A.8 \cite{BOG, Thm. IV.1.4}}
The weak topology on the non--degene\-rate indefinite inner product space $\VV$
is a majorant, iff $\dim \VV<\infty$.
\endlem
The indefinite inner product on a space  equipped with a majorant
Hilbert topology admits a simple description by the so--called {\it metric
operator}.
\prop{A.9 \cite{BOG, Thm. IV.5.2}}
Let $\VV$ be an indefinite inner product space with a majorant Hilbert
topology $\tau$ defined by a norm $||.||$. Then there exists a
Hermitean linear operator, called {\rm metric} (or {\rm Gram}) operator,
$G$ on $\VV$ such that 
$$
\ip x y = \sp x {Gy},\quad \forall x,y\in\VV,
$$
where $\sp . .$ is the positive inner product on $\VV$ that defines
$||.||$. Moreover, in this case $\VV$ is decomposable and the fundamental
decomposition can be chosen so that each of the three components is 
$\tau$--closed.
\endprop
The spaces we want to construct in the next section should be --- in a sense
--- complete.
\defi{A.10}
If a non--degenerate indefinite inner product space $\KK$ admits a decomposition
$$
\KK=\KK^+\os\KK^-,\quad \KK^+\subset\KK^{++},\ \KK^-\subset\KK^{--},
$$
where $\KK^+$, $\KK^-$ are complete respectively w.r.t. the restriction of
the weak topology to them (termed {\it intrinsically complete}),
then $\KK$ is called a {\it Krein space}. If moreover 
the rank of indefiniteness
of $\KK$ is finite, $\varkappa(\KK)<\infty$, then $\KK$ is called a
{\it Pontrjagin space}.
\enddefi
Krein spaces can easily be characterized:\nopagebreak
\prop{A.11 \cite{BOG, Thm. V.1.3}}
An indefinite inner product space $\VV$ is a Krein space iff
\roster
  \item "i)" There exists a majorant Hilbert topology $\tau$ on $\VV$ and
  \item "ii)" The metric operator $G$ is completely invertible.
\endroster
\endprop
The Hilbert--space--completion $\HH$ of an indefinite inner 
product space $\VV$, if 
it exists together with its metric operator $G$, is called
the {\it Hilbert space structure} $(\HH,G)$ associated to $\VV$.
In applications one would like to find the largest Hilbert space associated to 
an indefinite inner product space. For that, one considers {\it minimal}
majorant topologies, i.e., topologies $\tau_\ast$ such that no majorant
$\tau$ is weaker than $\tau_\ast$. Hilbert space structures given by
the completion of $\VV$ w.r.t. a minimal majorant are correspondingly
called {\it maximal}. We find that the Hilbert space structure is maximal,
iff it leads actually to a Krein space:
\lem{A.12 \cite{1, App. A.1}} 
A majorant Hilbert topology leads to a maximal Hilbert space structure
$(\KK,G)$, iff $G$ has a bounded inverse. Given  a Hilbert space structure
one can always construct a maximal one.
\endlem
The last statement means that every space admitting some majorant Hil\-bert 
topology can be completed to a Krein space.
%
\subhead The Geometry of Quasi--Positive Spaces \endsubhead
From now on we assume $\VV$ to be an infinite--dimensional,
non--degenerate, quasi--positive inner 
product space with rank of negativity $\nr(\VV)=N$, $N\in\bold N$. The
following remark sets up our framework.
\rem{A.13}
Let $\VV$ be as stated above. Equivalent are:
\roster 
  \item "i)" $\nr(\VV)=N$, $N\in\bold N$.
  \item "ii)" There exists a maximal neutral subspace $\VN\subset\VV^0$ with 
$\dim\VN=N$ such that $\VV=\Vp\dotplus\VN$, $\Vp\subset\Vpp$.
  Furthermore exists a maximal orthogonal system of $N$ neutral vectors
 that spans $\VN$: 
$\VN=\lin{\chi_1,\ldots,\chi_N}$, $\ip{\chi_i}{\chi_j} = 0,\ \forall 
i,j\in \{1,\ldots,N\}$.
\endroster
\endrem
{\proof{}
Assume that i) holds. Then Lemma A.4 shows that $\VV$ admits a fundamental
decomposition
$\VV=\Vp\os\Vm,\ \Vp\subset\Vpp,\ \Vm\subset\Vmm$, and 
$\Vm=\lin{v^-_1,\ldots,v^-_N}$, where the $v_1^-,\ldots,v_N^-$ are
assumed to form an orthonormal Basis of $\Vm$. Moreover, since $\VV$ is
infinite--dimensional, we find a orthonormal system
$\{v_1^+,\ldots,v_N^+\}\subset\Vp$ of positive vectors. Then the subspaces
$\VV_i\equiv\lin{v_i^+,v_i^-}$ are by construction linearly independent
and mutually orthogonal, and each $\VV_i$ contains a nonzero neutral
vector $\chi_i$, due to Lemma A.1. These vectors are an orthogonal
system spanning a subspace $\VN$. $\VN$ is neutral, since the $\chi_i$
are mutually orthogonal\footnote{Note that this construction would also
apply under the weaker assumption $\varkappa^+(\VV)\geq N$ instead of
$\dim \VV=\infty$, but not for general indefinite inner product
spaces. A counterexample would be $2+1$--dimensional spacetime, which
does not contain a neutral subspace of dimension $2$. I am
indebted to G. Hofmann for bringing this point to my attention.}.
We show that $\VN$ is maximal: Choose $x_i\in\VV$, 
$\ip{x_i}{\chi_i}\neq 0$,
which is possible since $\VV$ is non--degenerate. Then (Lemma A.2) 
$\lin{x_i,\chi_i}$ contains 
a negative vector $w_i^-$ and these vectors are again linearly independent.
Since by assumption the maximal number of linearly independent negative vectors
in $\VV$ is $N$, this shows that $\VN$ is maximal.
The other direction follows easily from Lemma A.2.
\ \hfill\ \qed
\endproof}
$\VV$ admits a majorant Hilbert topology, as we will now show. First note that
every $x\in\VV$ admits an unique decomposition
$$
x=x^+ + \IS x^i\chi_i,\quad x^+\in\Vp,\ x^i\in\bold C.\aeqno{1}
$$
\prop{A.14}
The seminorm 
$$
p(x)^2\equiv\ip{x^+}{x^+}+\IS\left\{|\ip{x}{\chi_i}|^2+|x^i|^2\right\},\quad
\forall x\in\VV, \aeqno{2}
$$
defines a majorant topology $\tau$ on $\VV$.  
\endprop
{\proof{}
Using the decomposition \aeqref{1} 
and $\ip{x^+}{\chi}=\ip{x}{\chi},\ \forall x\in\VV,\quad \chi\in\VN$,
the inner product on $\VV$ takes the form
$$
\ip{x}{y}=\ip{x^+}{y^+}+
\IS\left\{\cc{x^i}\ip{\chi_i}{y}+\ip{x}{\chi_i}y^i\right\},\quad 
\forall x,y\in\VV.
$$
Now the seminorm $p$ majorizes the inner square $\ip{x}{x}$, i.e.,
$|\ip{x}{x}|\leq p(x)^2,\ \forall x\in\VV$, since in every term of
the sum the estimate $|\cc{x^i}\ip{\chi_i}{x}+\ip{x}{\chi_i}x^i|\leq
|\ip{\chi_i}{x}|^2+|x^i|^2$ holds. Then Lemma A.7ii) shows that
$\ip{.}{.}$ is jointly $\tau$--continuous. Therefore $\tau$ is 
a majorant topology on $\VV$.
\ \hfill\ \qed
\endproof}
\cor{A.15}
On the closure $\KK\equiv\clot{\VV}$ of $\VV$ w.r.t. $\tau$ we can define
a Hilbert scalar product by
$$\sp{x}{y}\equiv\ip{x^+}{y^+}+\IS\left\{
\ip{x}{\chi_i}\ip{\chi_i}{y}+\cc{x^i}y^i\right\},\ 
\forall x,y\in\KK.\aeqno{3}$$
We denote the Hilbert norm on $\KK$ by $||.||\equiv p(.)$.
\endcor
{\proof{}
$\sp{.}{.}$ is well defined on whole $\KK$, since
$\ip{\ldotp}{\ldotp}$ has an unique extension to $\KK$ 
(denoted by the same symbol) by its joint 
continuouity w.r.t. $\tau$.
\ \hfill\ \qed
\endproof}
The following lemma sheds a first light on the structure of $\KK$.
\lem{A.16}
The subspace $\clot{\Vp}$ of $\KK$ is orthogonal to $\VN$ w.r.t. $\sp{.}{.}$.
That is, $\KK=\clot{\Vp}\oplus\VN$.
\endlem
{\proof{}
For $x^+\in\Vp$ and $\chi\in\VN$ \aeqref{3} reduces to 
$
\sp{x^+}{\chi}=\IS\ip{x^+}{\chi_i}\ip{\chi_i}{\chi}=0,
$
because $x^+$ and $\chi$ have decompositions with $(x^+)^i=0$ and $\chi^+=0$
respectively, and since all inner products vanish in the neutral subspace
$\VN$. Continuouity of $\ip{.}{.}$ then implies the statement.
\ \hfill\ \qed
\endproof}
To take a closer look on $\KK$ we consider the linear functionals
$$F_i(x)\equiv\ip{\chi_i}{x}$$
on $\VV$.
These functionals are nonzero since 
$\VV$ is non--degenerate, vanish on $\VN$, and are clearly bounded w.r.t.\ the 
seminorm $p$. Then, by the Hahn--Banach Theorem, $F_i$ has an extension to
$\KK$ (denoted by the same symbol), which is also bounded, and by \aeqref{2} 
we have $0<||F_i||\leq 1$. From now on we assume the $F_i$ to be normalized,
$||F_i||=1$, e.g., by choosing a new Basis 
$\tilde{\chi_i}\equiv\chi_i/||F_i||$ in $\VN$.
We have
\lem{A.17}
The vectors $v_i\in\KK$ representing $F_i$ by
$F_i(x)=\sp{v_i}{x}$, $\forall x\in\KK$
are actually in the closure of $\Vp$: $v_i\in\clot{\Vp}$.
\endlem
{\proof{}
$v_i$ exist in $\KK$ due to the Riesz Representation Theorem. Assume that
$(\vn)_{n\in\bold N}\subset\KK$ is a sequence converging to $v_i$, i.e.,
$||v_i-\vn||^2\longrightarrow 0$. Then, using decomposition \aeqref{1}
for $\vn$ (and adopting Einstein's summation convention), we find
$$\eqalign{
||v_i&-\vn||^2=
  \sp{v_i-(\vn)^+-(\vn)^j\chi_j}{v_i-(\vn)^+-(\vn)^k\chi_k}\cr
                &=
   ||v_i-(\vn)^+||^2-
   \sp{v_i-(\vn)^+-(\vn)^j\chi_j}{(\vn)^k\chi_k}-
   \sp{(\vn)^j\chi_j}{v_i-(\vn)^+}\cr
                &=
   ||v_i-(\vn)^+||^2-
   \sp{v_i-\vn}{(\vn)^j\chi_j}-
   \sp{(\vn)^j\chi_j}{v_i}+\sp{(\vn)^j\chi_j}{(\vn)^+}.}$$
The last term on the right hand side vanishes identically, due to Lemma A.16. 
The third term is 
zero, since $\sp{v_i}{\chi_j}=F_i(\chi_j)=0$. The second term is
bounded by
$|\sp{v_i-\vn}{(\vn)^j\chi_j}|$ $\leq||v_i-\vn||\JS|(\vn)^j|$.
In this expression the sum stays bounded, whereas the first factor tends to
zero for $n\rightarrow\infty$.
Hence necessarily $||v_i-(\vn)^+||\longrightarrow 0$, which proves
the statement.
\ \hfill\ \qed
\endproof}
Using this result we can collect some properties of the $v_i$:
\lem{A.18} It holds
\roster
  \item "i)" $\sp{v_i}{v_j}=\ip{\chi_i}{v_j}=0$, for $i\neq j$, and
             $||v_i||^2=\sp{v_i}{v_i}=\ip{\chi_i}{v_i}=1$.
  \item "ii)" $\ip{v_i}{v_i}=0$.
  \item "iii)" $\ip{v_i}{x}=x^i,\ \forall x\in\VV$.
\endroster
\endlem
{\proof{}
The second part of i) is clear.
Assume $\vn$ to be a sequence in $\Vp$ converging to $v_i$ in $\KK$.
Then with \aeqref{3}
$$
{\left({|\ip{\chi_i}{\vn}|^2}\over{||\vn||^2}\right)^{-1}}=
{{\ip{\vn}{\vn}^{\ }}\over{|\ip{\chi_i}{\vn}|^2}} +1+
  {\sum\limits_{j\neq i}{|\ip{\chi_j}{\vn}|^2}\over{|\ip{\chi_i}{\vn}|^2}}.
$$
Now $|\ip{\chi_i}{\vn}|^2=|\sp{v_i}{\vn}|^2\longrightarrow 1$, 
since $||v_i||^2=1$,
so that the left hand side tends to one for $n\rightarrow\infty$. 
By the same argument the denominators on the right hand side stay bounded.
Thus necessarily $|\ip{\chi_j}{\vn}|^2\longrightarrow 0$, showing the
first part of i). Furthermore $\ip{\vn}{\vn}\longrightarrow 0$, which shows ii),
since $\vn$ converges  w.r.t. $||.||=p(.)$ and $p(x)^2$ majorizes 
$|\ip x x |$.
To show iii), we consider again the decomposition \aeqref{1} for a vector
$x\in\VV$ which yields
$$
\ip{\vn}{x}=\ip{\vn}{x^+}+x^i\ip{\vn}{\chi_i}+
  \sum_{j\neq i}x^j\ip{\vn}{\chi_j}.
$$
In this expression $\ip{\vn}{x^+}\longrightarrow 0$, since by ii) $\vn$ 
converges strongly to zero in $\clot{\Vp}$. The sum tends to zero due to i), 
leaving us with the second term, which tends to $x^i$ for $n\rightarrow \infty$
by i), as proposed.
\ \hfill\ \qed
\endproof}
So the $v_i$ form an orthonormal system in $\KK$.
It is clear by now that the space $\lin{v_i|i=1,\ldots,N}\subset\KK$
is isomorphic to the dual space $\VND$ of $\VN$ w.r.t. $\ip{.}{.}$. We are now
ready to state our main result, which says roughly 
that every quasi--positive space
admits a maximal Hilbert space completion to a Pontrjagin space.
Note that this result has already been found by G. Hofmann,
cf. \cite{9}, in a different context, but without explicit construction
of the minimal majorant topology.
\theo{A.19}
The space $\KK$ is a Pontrjagin space with $\nr(\KK)=N$. Its Hilbert space
structure is maximal and given as follows:
\roster
  \item "a)" If $\HH$ is the Hilbert 
        space closure of $\Vp$ w.r.t. the topology $\tau_+$ induced by the 
        Norm $p_+(x^+)^2\equiv\ip{x^+}{x^+},\ \forall x^+\in\Vp$,
        then $\KK$ admits the orthogonal decomposition
        $$\KK=\HH\oplus\VND\oplus\VN.$$
  \item "b)" The metric operator $G$ on $\KK$, $\ip{.}{.}=\sp{.}{G.}$ has 
the form
$$
G={\boldkey 1}_\HH\oplus 
\underbrace{J \oplus \ldots\oplus J}_{N\ \roman{times}},\quad 
      J=\left(\matrix 0 & 1 \cr 1 & 0 \endmatrix\right):\ 
\widehat{\VV}_i\rightarrow\widehat{\VV}_i,
$$
where $\widehat{\VV}_i=\lin{v_i}\oplus\lin{\chi_i}$, $i=1,\ldots,N$.
\endroster
\endtheo
{\proof{}
To prove a), we have to show that 
$\clot{\Vp}=\HH\oplus\lin{v_i|i=1,\ldots,N}$, taking Lemma A.16 into account.
First, note that according to Lemma A.18 the $v_i$  form an orthonormal
basis for $\VND$. For $x^+, y^+\in\Vp$ we have by definition of the
$v_i$ and \aeqref{3}
$$
\sp{x^+}{y^+}=
\ip{x^+}{y^+}+\IS\sp{x^+}{v_i}\sp{v_i}{y^+}.
$$
This shows that a sequence $(x^+_n)_{n\in \boldkey N}\subset\Vp$ converges
to $x\in\KK$, iff $p_+(x_n^+-x)\longrightarrow 0$ and independently
the orthogonal projections of $x_n^+-x$ onto $\VND$ converge to zero.
This means that $\Vp\cap(\VND)^\bot$ is dense in $\HH$ w.r.t. $\tau_+$,
which shows that the decomposition of $\clot{\Vp}$ is indeed orthogonal.

If we can render the metric operator as stated in b), it follows that
the negative subspace of $\KK$ has dimension $N$, since every 
$\widehat{\VV}_i$
contains exactly one one--dimensional negative subspace, which is clear
from the action of $J$ on these spaces.
Then, since this metric operator is clearly invertible on whole $\KK$,
Proposition A.11 shows that the components of $\KK$ are intrinsically
complete, and therefore $\KK$ is a Krein space and actually a
Pontrjagin space, after Definition A.10.
Lemma A.12 then tells us that $(\KK, G)$ is maximal.

First for $x^\bot,y^\bot\in \Vp\cap(\VND)^\bot$ we have 
$\sp{x^\bot}{y^\bot}=\ip{x^\bot}{y^\bot}$. Thus $G$ acts as the identity
on this dense set in $\HH$, and therefore also on whole $\HH$.
Further by definition $\ip{x}{\chi_i}=\sp{x}{G\chi_i}=\sp{x}{v_i},\ 
\forall x\in\KK$, showing that $G\chi_i=v_i$.
On the other hand $\ip{x}{v_i}=\cc{x^i}=\sp{x}{\chi_i},\ \forall x\in\KK$ 
by Lemma A.18iii) and \aeqref{3}. This shows $Gv_i=\chi_i$, which yields
the desired result.
\ \hfill\ \qed
\endproof}
%
\subhead Application: Integral Kernels with Algebraic Singularities \endsubhead
We consider the space $\SR$ of smooth, complex valued test functions, 
decreasing faster (together with all derivatives) than $1/|x|^n$ for
all $n\in\bold N$. We assume $\Kl$ to be a {\it positive integral Kernel with
algebraic singularity} of strength $\lambda\in\bold C$ at $0$. By this we mean,
that $\Kl$ defines a linear functional $\Kl$ on $\SR$ by the {\it canonical 
regularization} (cf. \cite{GS}, \S I.3), which is given for all $f\in\SR$ by
$$
 \Kl(f)\equiv\kint
   \left\{f(x)-\ISi{x^i \over i!}f^{(i)}(0)\right\},
\aeqno{4}
$$
where $-N-1<\RE \lambda<N$, $N\in\bold N$,
and $f^{(i)}$ denotes the $i$--th derivative of $f$, $f^{(0)}\equiv f$. 
We make $\SR$ an inner product 
space by introducing the sesquilinear form
$$
\ip f g \equiv \Kl(\cc f g),\ \forall f,g\in\SR.
\aeqno{5}
$$
This is clearly an indefinite inner product on $\SR$, the indefiniteness
coming from the $N$ regularization terms at the origin.
For simplicity we assume this inner product  to be non--degenerate
(otherwise one would just have to go over to equivalence classes in $\SR$).
Furthermore, by the term {\it positive} kernel we mean that $\ip . .$ shall
be a positive form, or equivalently that $\Kl$ is a positive measure 
on the subspace $\SRp\subset\SR$, which will be defined below.

For the sake of concreteness, consider the real functions
$\tilde{\chi}_i(x)\equiv x^{i/2}\vartheta_i(x)\in\SR$ for $i=0,\ldots,N-1$,
where $\vartheta_i\in\SR$ is a function which is unity in a neighbourhood
of the origin. The $\vartheta_i$ shall now be chosen such that the condition
$\ip{\tilde{\chi}_i}{\tilde{\chi}_i} = 0$ holds, i.e., such that the
$\tilde{\chi}_i$ are neutral vectors. 
Also, $\vartheta_i$ must be chosen such that we can obtain
an orthogonal system of neutral vectors $\chi_i$, which span the
same subspace as the $\tilde{\chi}_i$:
$$\eqalign{ &
\VN\equiv\lin{\chi_i|i=0,\ldots,N-1}\equiv\lin{\tilde{\chi}_i|i=0,\ldots,N-1}
  \subset\SR,\cr
 & \ip{\chi_i}{\chi_j} = 0,\quad \forall i,j\in\{0,\ldots,N-1\}.\cr
}
$$
A concrete construction of such a function system can be found 
in~\cite{10}, Appendix~A.
Then every $f\in\SR$ admits a linear decomposition
$$
f=f^++\ISi f^{(i)}(0)\chi_i,\quad f^+\in\SRp.
\aeqno{6}
$$
The remaining part $f^+$ of $f$ is then in the subspace $\SRp\subset\SR$
of functions for which the unregularized Integral
$$
\ip{f^+}{f^+}\equiv\Kl(|f^+|^2)\equiv\kint\,|f^+(x)|^2\aeqno{7}
$$
is positive and finite. The positive definite scalar product
\aeqref{3} then takes the special form 
$$
\sp{f}{g}\equiv\ip{f^+}{g^+}+\ISi\left\{
\ip{f}{\chi_i}\ip{\chi_i}{g}+\cc{f^{(i)}(0)}g^{(i)}(0)\right\}
\aeqno{8}
$$
for all $f,g\in\SR$,
inducing a seminorm $p(f)^2\equiv\sp f f$, and by that a majorant
Hilbert topology $\tau$ on $\SR$.

We are now in a position to apply the results of the last section to
$\SR$. The main Theorem A.19 shows that $\clot{\SR}$ has the structure
$\HH\oplus\VND\oplus\VN$. Now the restriction of the norm $||f||^2\equiv
\sp f f$ to $\SRp$ is nothing but the $\LL$--norm for functions on $\bold R$,
measurable w.r.t. $dx\,\Kl$, showing that the Pontrjagin space with rank
of negativity $\nr(\clot{\SR})=N$ in this case is
$$
\clot{\SR}\equiv\LKL\oplus\VND\oplus\VN.
\aeqno{9}
$$
%
\head Acknowledgements \endhead
I would like to thank G. Hofmann for pointing out an error in a previous
version of the manuscript. Thanks to R. H\"aring and D. Lenz for
discussions, and to Daniel~A.~Dubin for pointing out references~\cite{11}
and~\cite{12}.
%
%

%
%
\medskip
{\flushpar
\eightpoint
\rec{{June 26, 1996; updated March 4, 2002}}
\addr{Andreas U.~Schmidt}
\addr{Fachbereich Mathematik}
\addr{Johann Wolfgang Goethe--Universit{\"a}t}
\addr{D--60054 Frankfurt am Main}
\addr{{\it{aschmidt\@math.uni--frankfurt.de}}}
\enddocument